\documentclass[prb,twocolumn,aps,notitlepage,
superscriptaddress]{revtex4-1}
\usepackage{graphicx}

\usepackage{xcolor}

\usepackage{hyperref}
\usepackage[utf8]{inputenc}
\usepackage[T1]{fontenc}
\usepackage{amsmath}

\DeclareMathOperator{\sign}{sign}

\newcommand*{\ket}[1]{\left|#1\right\rangle}
\newcommand*{\bra}[1]{\left\langle#1\right|}
\def\arttit{Topological Josephson junction in transverse magnetic field}
\begin{document}

\title{\arttit}

\author{Stefan \surname{Backens}}
\affiliation{Institut f\"ur Theorie der Kondensierten Materie, Karlsruhe Institute of Technology, D-76131 Karlsruhe, Germany}
\author{Alexander \surname{Shnirman}}
\affiliation{Institut f\"ur Theorie der Kondensierten Materie, Karlsruhe Institute of Technology, D-76131 Karlsruhe, Germany}
\affiliation{Institut f\"ur QuantenMaterialien und Technologien, Karlsruhe Institute of Technology, Hermann-von-Helmholtz-Platz 1, D-76344 Eggenstein-Leopoldshafen, Germany}
\author{Yuriy \surname{Makhlin}}
\affiliation{Condensed-matter physics Laboratory, HSE University, 101000 Moscow, Russia}
\affiliation{Landau Institute for Theoretical Physics, acad.~Semyonov av. 1a, 142432, Chernogolovka, Russia}

\begin{abstract}
We consider Majorana zero modes in a Josephson junction on top of a topological insulator in transverse magnetic field. 
Majorana zero modes reside at periodically located nodes of Josephson vortices. We find that hybridization of these modes is prohibited by symmetries of the problem at vanishing chemical potential, which ensures better protection of zero modes and yields methods to control the tunnel coupling between Majorana modes for quantum information processing applications.
\end{abstract}

\maketitle

Topologically-protected quantum manipulations with Majorana zero modes (MZM's) are extensively studied theoretically and experimentally due to their exotic properties, including exchange statistics, and their possible use in platforms for topological quantum computation~\cite{NayakReview08, DasSarmaNJP15, KitaevTopQC}.
In particular, hybrid superconductor-topological insulator structures were discussed. Fu and Kane analyzed a topological Josephson junction between superconductor films on top of a topological insulator~\cite{FuKane08,FuKane09} and demonstrated the appearance of Majorana edge states. Here we consider a setup where Majorana bound states are point-like structures bound to Josephson vortices in an external magnetic field perpendicular to the surface~\cite{PotterFu13}. Such devices were discussed as a platform for topological quantum computation with the possibility of manipulation and braiding of Majorana bound states~\cite{Vishveshwara}. We analyze the tunnel coupling between the MZM's and find that this coupling vanishes at zero chemical potential. This should be taken into account in the design of experiments with MZM's on Josephson vortices and also suggests that coupling and hybridization of various MZM's may be controlled, in particular, via the chemical potential. Note similar observations for a 2D vortex lattice~\cite{PikulinFranz}.

\begin{figure}[h]
\centerline{\includegraphics[width=0.35\textwidth]{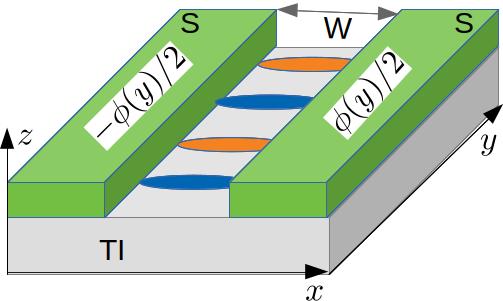}}
\caption{
S-TI-S Josephson junction in a transverse magnetic field $\parallel\hat z$. Blue and orange spots indicate location of Majorana bound states.}
\label{Fig:sample}
\end{figure}

\paragraph{Model Hamiltonian.}
We consider an S-TI-S Josephson junction between two s-wave superconducting (S) electrodes on top of a topological-insulator (TI) material, Fig.~\ref{Fig:sample}. Due to proximity effect, superconducting correlations are induced in the surface layer of the topological insulator. The states in this layer can be described by the Bogolyubov-de-Gennes Hamiltonian 
$H = \frac{1}{2} \int dx dy \,\Psi^\dag h \Psi$, where 
$\Psi = [\psi_\uparrow,\psi_\downarrow,\psi^\dag_\downarrow,-\psi^\dag_\uparrow]^T$.

The single particle BdG Hamiltonian reads
\begin{eqnarray}
h &=&  v\left(\sigma_x \,\left[- i\hbar\,\partial_x\right] + \sigma_y \left[- i\hbar\,\partial_y + \frac{e}{c} A_y(x)\tau_z\right]\right)\tau_z\nonumber\\
&+& \Delta(x,y)\tau_+ + \Delta^*(x,y)\tau_- - \mu\tau_z
\,,
\end{eqnarray}
where the Pauli matrices $\sigma$ and $\tau$ refer to the spin degree of freedom and to the Bogolyubov-Nambu particle-hole space ($\tau_\pm=(\tau_x\pm i\tau_y)/2$), respectively.

The vector potential $\bf A$ describes the distribution of the transverse magnetic field ${\bf H}=H\hat z$ around the gap of width $W$ between the superconducting electrodes:
\begin{equation}
H = \left\{\begin{matrix}
H_0 \exp\left[-\frac{|x|-W/2}{\lambda_L}\right] \,,
& |x|>W/2\,,\\
H_0\,, & -W/2<x<W/2\,.
\end{matrix}\right. 
\end{equation}
We choose the gauge with $\Delta=\Delta_0 e^{i\phi(y)/2}$ in one superconductor, $x>W/2$, and $\Delta=\Delta_0 e^{-i\phi(y)/2}$ for $x<-W/2$, while $\Delta=0$ in the gap $|x|<W/2$. The vector potential in this gauge has direction $y$ and is given by
\begin{eqnarray}
A_y(x) = \left\{\begin{matrix}
-H_0 x \,,&|x|<W/2\,,\\
\lambda_L H_0
e^{-\frac{x-W/2}{\lambda_L}} -
\frac{2\lambda_L +W}{2}\,H_0\,,&x>W/2\,,\\
-A_y(-x)\,,&x<0 \,.
\end{matrix}\right.
\end{eqnarray}

The phase is chosen as $\phi(y) = (2\lambda_L+W)\, H_0\, \frac{2\pi}{\Phi_0}\,y$ with $\Phi_0=hc/2e$ being the superconducting flux quantum. We assume a not too strong field so that the corresponding magnetic length $l_B = [(2\lambda_L+W)\, H_0/\Phi_0]^{-1}$ exceeds the relevant coherence length $\xi=\hbar v/\Delta_0$, $l_B\gg\xi$.

Although in a realistic setup the system may be in the Pearl regime~\cite{Pearl} with two-dimensional magnetostatics, we consider here the London regime~\cite{Tinkham}. While realistic situations may require further analysis, we expect on symmetry grounds that its basic properties will be the same.

\paragraph{Symmetries.}
Properties of the solutions are to large extent determined by symmetries of the Hamiltonian, which we discuss here. First, as usual for the BdG Hamiltonian, the charge conjugation or particle-hole symmetry $C = \sigma_y \tau_y K$ inverts energies: $C^{-1} h C = - h$.

Further, for  $\mu=0$, the case of our special interest below, in addition there is a (quasi) time-reversal symmetry
$T = \sigma_x \tau_x K$, which commutes with the Hamiltonian, $T^{-1} h T =  H$, and $T^2=1$. Thus, for $\mu=0$, as usual, the product of $C$ and $T$ is the chiral symmetry
$S=\sigma_z \tau_z$.
This set of symmetries describes the BDI symmetry class~\cite{AltlandZirnbauer,Schnyder08}.

Finally, we note that there is an extra symmetry operator,
\begin{equation}
F = \sigma_x \tau_x I_x \,,
\end{equation}
where the operator of $x$ inversion acts as 
$I_x \psi(x,y) = \psi(-x,y)$, $I^{-1}_x A_y(x) I_x = A_y(-x)$.

\paragraph{1D slices.}
We split the Hamiltonian into $h=h_0 + h_1$, where
\begin{equation}
h_0(y) = - i\hbar v\,\sigma_x \,\tau_z\partial_x  + [\Delta(x,y)\tau_+ + \textrm{h.c.}] -\mu\tau_z
\end{equation}
and
\begin{equation}
h_1 = v\, \sigma_y \left[- i\hbar\partial_y + \frac{e}{c} A_y(x)\tau_z\right]\tau_z \,.
\end{equation}

The Hamiltonian $h_0$ depends on $y$ as a parameter via $\Delta$ since its phase $\phi$ is proportional to $y$. Hence we can diagonalize it separately for each $y$ and then take into account $h_1$, which glues these 1D states into 2D wave functions. 
Thus we look for eigensolutions of
\begin{equation}\label{eq:h0eigen}
\hat h_0(y) \ket{\nu}_y = \epsilon_\nu(y) \ket{\nu}_y
\end{equation}
or equivalently, $\hat h_0(\phi) \ket{\nu}_\phi = \epsilon_\nu(\phi) \ket{\nu}_\phi$, where $y$ is a parameter and $\ket{\nu}_y$ represents a wave function $f_{\nu y}(x)$.

The resulting energies depend on system parameters, with examples shown in Fig.~\ref{Fig:BdGLevelsW=2,5} and  
Fig.~\ref{Fig:BdGLevelsW=6,0}.
\begin{figure}[h]
\centerline{\includegraphics[width=0.35\textwidth]{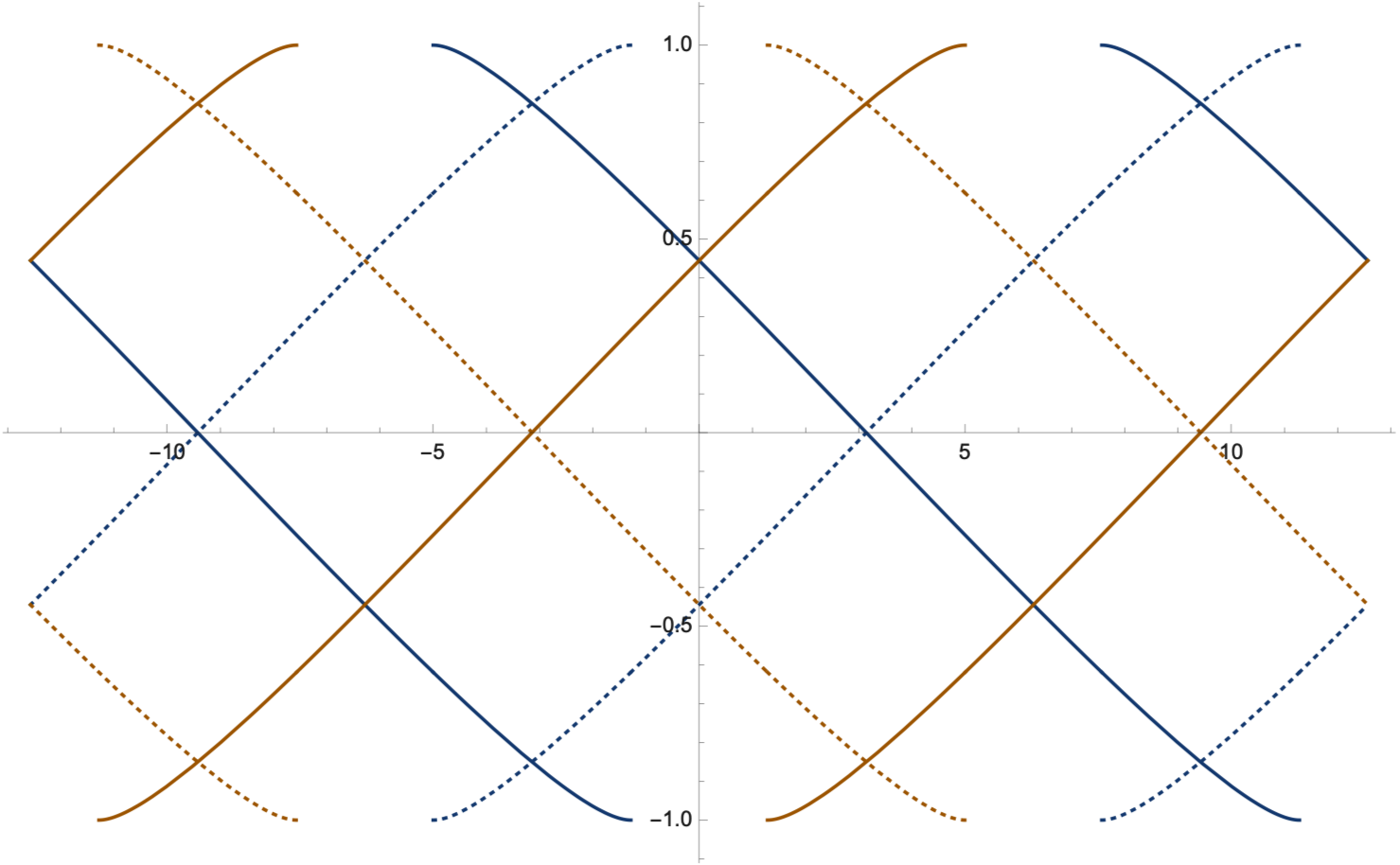}}
\caption{
Eigenenergies $\epsilon_n(\phi)$ of $h_0$ for $W=2.5\xi$}
\label{Fig:BdGLevelsW=2,5}
\end{figure}
\begin{figure}[h]
\centerline{\includegraphics[width=0.35\textwidth]{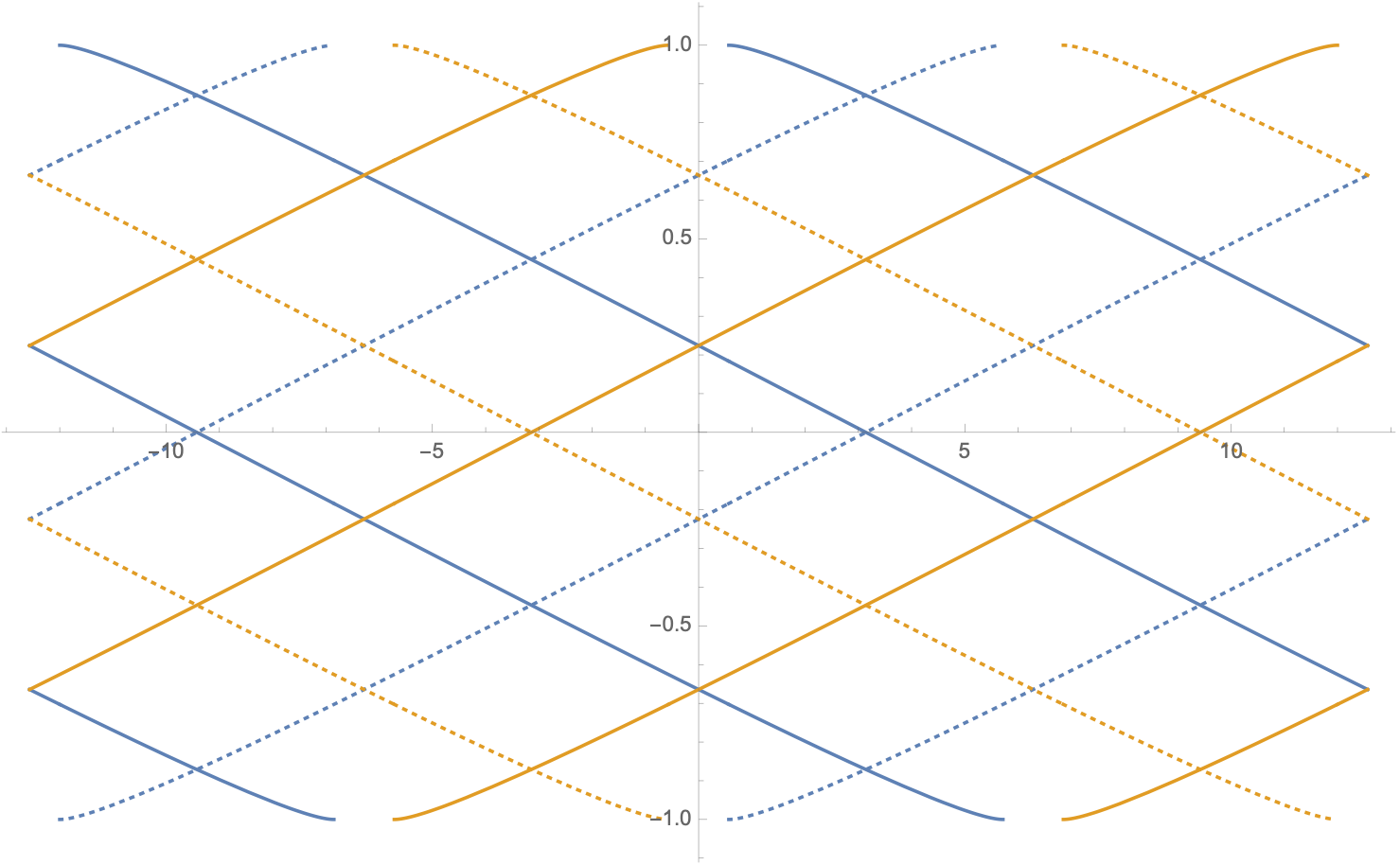}}
\caption{
Eigenenergies $\epsilon_n(\phi)$ of $h_0$ for $W=6.0\xi$}
\label{Fig:BdGLevelsW=6,0}
\end{figure}
The eigenstates can be classified by the eigenvalues $\pm 1$ of the  symmetry operator $F$ (as indicated with the blue, $F=1$, and orange, $F=-1$, color in the figures) and $\sigma_x$ (a symmetry of $h_0$, see below).

The spectrum $\epsilon_\nu$ depends on $\phi$ (or $y$), and 
has zero modes~\cite{FuKane08} at $\phi=\pi+2\pi n$: 
here the gap is purely imaginary $\Delta(x)=i(-1)^n\Delta_0\sign x$, and
\begin{equation}
h_0 = -iv\sigma_x\tau_z\partial_x
- (-1)^n \Delta_0 \tau_y \sign x
\end{equation}
has two zero modes, which are eigenstates of $F=\sigma_x\tau_x I_x$ with $F=(-1)^n$ ($F=1$ at $\phi=\dots-3\pi,\pi,5\pi\dots$ and $F=-1$ at $\phi=\dots-\pi,3\pi,7\pi\dots$):
\begin{equation}\label{eq:lowEn}
\begin{pmatrix}1\\\sigma_x\\\sigma_x F\\F\end{pmatrix}
\cdot e^{- \int_0^{|x|} |\Delta(x')|dx'}
\end{equation}
(note another gauge used in Ref.~\cite{FuKane08}).

It is convenient to mark an eigenstate $\epsilon_\nu(y)$ of $h_0$ with the sign of its slope ($\pm = - \sigma_x$) and an integer $n$, where its energy crosses zero, $\epsilon_\nu(\phi=\pi+2\pi n)=0$, so that $\nu=(n,\pm)$.

We observe that the spectrum is $2\pi$-periodic, $\epsilon_\nu(\phi+2\pi)=\epsilon_{\nu'}(\phi)$, 
however the wave functions are $4\pi$-periodic, $f_\nu(\phi+4\pi)=f_{\nu''}(\phi)$. This is another manifestation of the symmetry $F$. 

\paragraph{Construction of 2D wave functions.}
Having obtained the eigenvectors $\ket{\nu}_y$ (\ref{eq:h0eigen}) of $h_0$ for each $y$, we use them as a basis to construct solutions
\begin{equation}
\psi(x,y) = \sum_\nu \alpha_\nu(y) f_{\nu y}(x) \,,
\end{equation}
of the 2D BdG equation
\begin{equation}
(h_0+h_1)\psi = E\psi \,.
\end{equation}
Rewriting the BdG equation as $(E-h_0)\psi = h_1 \psi$,
we find
\begin{align}
(E - \epsilon_\lambda(y)) \alpha_\lambda(y)  = \sum_\nu
(h_\textrm{kin}^{\lambda\nu} \partial_y
+ \tilde h_1^{\lambda\nu}) \alpha_\nu(y) \,,
\label{eq:BdGforalphan}\\
h_\textrm{kin}^{\lambda\nu} =
(-i\hbar v)\, {}_y\!\bra{\lambda}\sigma_y\tau_z \ket{\nu}_y \,,
\\
\tilde h_1^{\lambda\nu} = (-i\hbar v)\, {}_y\!\bra{\lambda}\sigma_y\tau_z \left(\partial_y +\frac{ie}{\hbar c}A_y(x)\tau_z\right)\ket{\nu}_y \,.
\end{align}
We note that Fu and Kane~\cite{FuKane08} used a different, $y$-independent basis $\ket{\nu}_{y=y_\lambda}$, suitable at $y\approx y_\lambda$ for low-energy states near a node $\epsilon_\lambda(y_\lambda)=0$. Since we consider also ranges away from nodes, we chose $\ket{\nu}_y$ above. This brings a new term $\tilde h_1$ compared to Ref.~\cite{FuKane08}, which we will treat perturbatively.

\paragraph{Additional symmetries of $h_0$ at $\mu=0$.}
$h_0$ has an extra symmetry
\begin{equation}
[h_0,\sigma_x]=0 \,,
\end{equation}
which at $\mu=0$ allows one to use an alternative time reversal $\Theta = \tau_x K = \sigma_x T$ and an alternative chiral symmetry $P = \sigma_y \tau_z \propto \sigma_x S$.
Note that $\Theta$ and $P$ are not symmetries of the full Hamiltonian $h_0+ h_1$: indeed, $\{h_1,\Theta\}=0$ whereas $[h_0,\Theta]=0$, also $[h_1,P]=0$ whereas $\{h_0,P\}=0$.

The chiral symmetry $P$ maps between $\ket{n\pm}$ and squares to unity, $P^2=1$, so we may choose phase factors such that $P\ket{n+} = \ket{n-}$ and $P\ket{n-} = \ket{n+}$.

\paragraph{Zero modes.}

Let us first omit the $\tilde h_1$ term in Eq.~(\ref{eq:BdGforalphan}). As the symmetries above show, $h_\textrm{kin}$ couples only the levels within each charge conjugate pair $n\pm$, and Eq.~(\ref{eq:BdGforalphan}) factorizes into $2\times 2$ equations:
\begin{equation}
E \left(\begin{array}{c}\alpha_{n+}(y) \\ \alpha_{n-}(y) \end{array}\right) = h_{\rm eff} \left(\begin{array}{c}\alpha_{n+}(y) \\ \alpha_{n-}(y) \end{array}\right) \,,
\end{equation}
where
\begin{equation}
h_{\rm eff} = - i v \rho_y \partial_y + \epsilon_n(y) \rho_z \,.
\end{equation}
Here $\rho$ are a new set of Pauli matrices in the basis $n\pm$.

\begin{figure}[h]
\centerline{\includegraphics[width=0.2\textwidth]{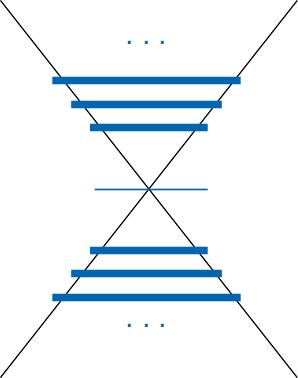}}
\caption{
Schematic drawing of energy levels near a single node.}
\label{Fig:specCone}
\end{figure}

The energy $\epsilon_n$ vanishes at coordinates $y=y_n$ with phase $\phi(y_n)=\pi+2\pi n$. Around this point we linearize $\epsilon_n(y) = \alpha\,(y-y_n)$ and obtain an exactly solvable~\cite{PotterFu13,Vishveshwara}
\begin{equation}\label{eq:1node}
h_{\rm eff} = - i v \rho_y \partial_y + \alpha\,(y-y_n) \rho_z
\end{equation}
with $\alpha\propto\partial_y\phi$.
This equation has a zero mode localized near $y_n$ and a set of `Landau levels', cf.~Fig.~\ref{Fig:specCone}. In the regime we  consider, $l_B\gg\xi$, many Landau levels fit below the gap $\Delta_0$.

\paragraph{Coupling and hybridization of Majorana zero modes.}

Taking into account the $\tilde h_1$ terms in Eq.~(\ref{eq:BdGforalphan}) may tunnel-couple zero modes at different nodes $\phi=\pi+2\pi n$, pushing them away from zero energy. We show that this is not the case, and at $\mu=0$ the modes remain at zero energy, forming a flat band. This observation is similar to that of Ref.~\cite{PikulinFranz} for a 2D system with a vortex lattice.

Indeed, from Eq.~(\ref{eq:1node}) one observes that the zero mode at each node, $\ket{n+}+S\ket{n-}$, is chiral with the same chirality $S=\sigma_z\tau_z=-\sign\alpha$ for all nodes. We show that adding $\tilde h_1$ does not alter this property.

We note first that effects of $\tilde h_1$ on the energies of the zero modes are weak and can be treated perturbatively, and hence the zero modes may acquire only a weak admixture of other, non-zero states. On the other hand, the chiral symmetry implies that all zero-energy solutions have full chirality, while all nonzero-energy solutions are equal-weight superpositions of two chiralities. Since without $\tilde h_1$ all zero modes have the same chiralities, a weak perturbation cannot push them away from zero.

For a sufficiently wide contact $W\gg\xi$ the spectrum $\epsilon_\nu(y)$ is below $\Delta_0$ and linear in a wide range of phases, so that the zero mode wave functions can be found explicitly in the range of interest. Then one observes the vanishing tunnel coupling explicitly. The zero states, for instance, in two neighboring orange nodes $\phi=-\pi$ and $\phi=3\pi$ are both $\propto(1,0,0,1)^T$, and the matrix element of the Hamiltonian between them vanishes, $\langle\psi_{-\pi}|h|\psi_{3\pi}\rangle =0$ as follows from its matrix structure (and is a consequence of the fact that $h$ changes chirality).

While the zero modes remain decoupled, other levels near a certain node can be perturbed due to coupling to other nodes. This is schematically indicated in Fig.~\ref{Fig:specCone} by a finite width of the corresponding Landau level.

Our result is in agreement with the general classification of zero modes~\cite{TeoKane}, which implies that the number of MZM's is topologically protected (note the same chirality of all MZM's) and given by the total phase drop accumulated around the defects within the Josephson junction.

\paragraph{Discussion.}
We note that coupling between the MZM's in this setting was suggested as a method to braid them~\cite{Vishveshwara}, which would allow for topologically-protected quantum operations~\cite{Alicea10,Alicea12}. While the coupling vanishes at $\mu=0$, a finite coupling may be achieved for $\mu\ne0$ (in fact, reaching $\mu=0$ may be experimentally challenging), which also permits controlling its strength.

Furthermore, finite coupling can be realized if two MZM's have different chiralities.
Since the chirality depends on the sign of the magnetic field ($\sign\alpha$ in Eq.~(\ref{eq:1node})), this can be effected with a nonuniform magnetic field distribution: for instance, with a 'domain wall' of $H(y)=H_0\sign y$ or an oscillatory $H(y)$. The corresponding structures would allow for controlled quantum operations with Majorana modes.

This research was supported by the DFG under grants No.~SH81/6-1 and  No.~SH81/7-1 and RSF under No.~21-42-04410.

\bibliography{longjj}

\end{document}